Defect model for the mixed mobile ion effect revisited: an importance of deformation rates

Vladimir Belostotsky

1005 Curtis Place, Rockville, MD 20852

Email: vladbel@erols.com

ABSTRACT

The progress in understanding the behavior of glassy mixed ionic conductors within the concept of the defect model for the mixed mobile ion effect (V. Belostotsky, J. Non-Cryst. Solids 353 (2007) 1078) is reported. It is shown that in a mixed ionic conductor (e.g., mixed alkali glass) containing two or more types of dissimilar mobile ions of unequal size sufficient local strain arising from the size mismatch of a mobile ion entering a foreign site can not be, in principle, absorbed by the surrounding network-forming matrix without its damage. Primary site rearrangement occurs immediately, on the time scale close to that of the ion migration process, through the formation of intrinsic defects in the nearest glass network. Neither anelastic relaxation below glass transition temperature, $T_g$, nor viscoelastic or viscous behavior at or above $T_g$ can be expected being observed in this case because the character of the stress relaxation in a wide temperature range is dictated above all by the deformation rates employed locally to the adjacent network-forming matrix. Since the ion migration occurs on the picosecond time scale, the primary rearrangement of the glass network adjacent to an ionic site occurs at rates orders of magnitude higher than those of the critical minimum values, so the matrix demonstrates brittle-elastic response to the arising strain even at temperatures well above $T_g$, which explains, among other things, why mixed alkali effect is observable in glass melts.

1. INTRODUCTION

Recently introduced defect model for the mixed mobile ion effect (MMIE) [1] is based upon the notion that simultaneous migration of two or more dissimilar mobile ions of unequal size in mixed alkali glasses causes the formation of intrinsic structural defects in glass network in the nearest vicinity of the ionic sites where foreign ions replace host ones. These defects, in oxide glasses these are primarily non-bridging oxygens (NBOs), acting as chain-terminators interrupt the covalent backbone, reduce its connectivity and increase its fragmentation which cause a reduction of glass viscosity [2,3]. Since transition temperature of glass is strictly linked to the density of the network cross-linking [2], $T_g$ of mixed alkali glasses is observed at lower temperatures than expected from the linear interpolation of the transition temperatures of the corresponding single alkali glasses exhibiting a minimum at the midpoint of mixing. In addition, generation of intrinsic defects as a result of the rearrangement of the sites where foreign ions replace host ones leads to the formation of sites where alkali cations directly interact with more than one NBO. These 'non-equilibrium' alkali sites act as high-energy anion traps in glass network causing orders of magnitude decrease in ionic conductivity with minimum where dissimilar alkali concentrations are close to each other. As was demonstrated, the defect model provides a unified, comprehensive and generally applicable fundamental explanation for the mixed alkali effect in glass in all its features, and agrees, at least qualitatively, with all experimental facts.

The central premise underlying the defect model is that the site rearrangement and relaxation of the primary stress arising from the size mismatch of a foreign ion replacing a host one in ionic site shall occur immediately, on the time scale close to that of the ion migration process. In response to applied stress chemical bonds in glass network adjacent to ionic site behave as ideal springs and snap when the deformation exceeds the bond-breaking threshold. In other words, network-forming matrix is presumed to demonstrate brittle-elastic response to the applied stress in the nearest vicinity of the ionic sites where foreign ions of different size replace host ones, and this type of behavior was suggested to be common as to glass below $T_g$ as to glass melt.

In [1] the ideas described above were assumed being intuitively obvious and as such were not conferred at that time. So, the primary purpose of the work reported here is to give an account on this matter.

## 2. THE NATURE OF THE NETWORK-FORMING MATRIX RESPONSE TO THE DISSIMILAR ION MOVEMENT

A generally accepted viewpoint among glass scientists holds that below $T_g$ glass network should exhibit an anelastic response to mechanical stress due to the presence of residual looseness in glassy phase where a local displacement of atomic particles at loose spots shall serve to relax the applied stress in the jammed structure in a reversible way, whereas at and above $T_g$ a glassy system is expected to demonstrate viscoelastic or viscous response to the applied stress [4-8]. These structural relaxation processes are thought to occur on the time scale much longer than those of the ion migration processes at low and intermediate temperatures. Only at very high temperatures close to liquidus characteristic times of structural relaxation and ion migration become compatible [4].

From the first glance, this viewpoint contradicts to what is proposed in the defect model for the MMIE. The resolution of the disagreement requires a more detailed analysis of the behavior of glass network under stress.

Obviously, when a mobile ion jumps into an open and available site previously occupied by an ion of different size, the site must be reconfigured immediately upon the ion arrival. An ion simply can not 'stay' at a doorway and 'wait' for, say, 100 s at $T_g$ or even for $10^{-2}$ s well above $T_g$ until a site would 'decide' to adjust itself to allow an ion of different size to enter. Attaining sufficient kinetic energy, a mobile ion either jumps directly into a vacant site, or rolls back. Irrespective of temperature, glass network in the nearest vicinity of an ionic site is forced to respond to the arrival of a foreign ion on the same time scale as an ion migration process.

The reason why site rearrangement has to proceed through the adjacent network bond breakup and defect generation below $T_g$ is quite obvious: At low temperatures silicate glass is well known to be a truly brittle material behaving ideally elastic under mechanical stress. There is no plastic yielding, the deformation energy is entirely stored as elastic strain, and all stored potential energy dissipates instantaneously through the creation and propagation of cracks when the material reaches the damage threshold. From the microscopic viewpoint this means that below $T_g$ the bonds between the network-forming atoms behave as perfect linear springs until the instant they snap when the distance between the atoms exceeds the bond-breaking threshold.

TABLE I. Deformation rates ($\dot{\varepsilon}$) of the first coordination shells for alkali cation pairs in mixed alkali silicate glasses

| Ionic pair | Li-Na | Na-K | K-Cs |
|---|---|---|---|
| Deformation Rate, m·s$^{-1}$ | 270 | 300 | 190 |

The legitimacy of viewing the character of the response of glass matrix to stress arising from the size mismatch of a mobile ion in a foreign site in mixed alkali glasses as brittle-elastic at and above $T_g$ is not so evident from the first glance. However, it can barely be questioned when we turn to the works of Brückner and co-workers [9,10] who have studied the brittleness of glass melt at high deformation rates. Their findings show that for each given temperature/viscosity a critical minimum deformation rate exists above which glass melt demonstrates brittle-elastic response to the applied stress.

In mixed alkali glass the deformation rate, $\dot{\varepsilon}$, employed locally to the glass network in the nearest vicinity of an ionic site where a foreign ion of different size replaces a host one can be defined as a ratio of elongation or shrinkage of a circumference of the first coordination shell of the ionic site, $\Delta L_{FCS}$ (see Tab.II in the Ref. 1), to the characteristic time scale describing the ion migration process, $\tau_O$, which is of order of $10^{-12}$ s:

$$\dot{\varepsilon} = \Delta L_{FCS} / \tau_O$$

Deformation rates calculated for several alkali pairs are shown in the Tab. I.

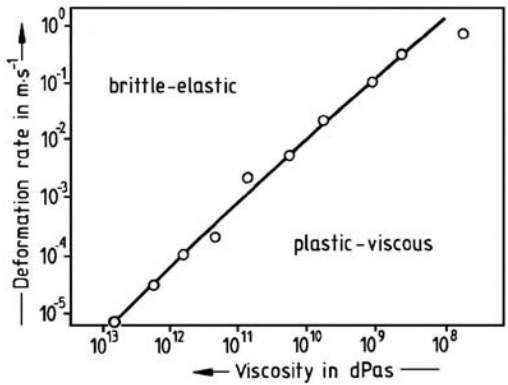

Fig. 1. Critical minimum deformation rate vs. viscosity for brittle fracture of silicate glass melt. Glass transition temperature, $T_g$, corresponds to the viscosity of $10^{13}$ dPas (Redrawn from Ref. 10)

A comparison of the calculated values and critical minimum deformation rates for brittle fracture of silicate glass melt experimentally obtained by Brückner and co-workers (Fig. 1) [10] indicates that at $T_g$ local deformation of glass network occurs at a rate more than seven orders of magnitude higher than the critical minimum value, and it is at least two orders of magnitude higher as melt viscosity reaches $10^8$ dPas within the glass transformation region. This makes it immediately obvious that viscoelastic or viscous response in this case can not be expected because extremely high-rate deformation is employed so that even well above $T_g$ glass-forming melt behaves totally unrelaxed like a brittle body with exclusively elastic components [10]. This explains why mixed alkali effect is observed not only in glasses below $T_g$ but also in glass-forming melts.

As was put forward in [1], further structural rearrangement may occur in glass matrix which involves much slower diffusion-driven annealing of newly created intrinsic defects below $T_g$ or/and reformation of the basic Si-O-Si linkages via thermally activated shedding and recoupling above $T_g$. Thus, two relaxation processes take place: rapid forced site rearrangement on the picosecond time scale which results in the formation of intrinsic defects in glass matrix adjacent to the sites, and subsequent much slower network relaxation. These processes become increasingly more and more competing with rising the temperature, so that the magnitude of the MMIE is defined by the interplay between kinetics of the structural defect generation and relaxation.

CONCLUSION

It is shown that in mixed alkali glasses the local deformation of glass network in the nearest vicinity of ionic sites where foreign ions replace host ones occurs at rates orders of magnitude higher than the critical minimum deformation rates for the brittle fracture of glass in wide temperature range including glass transition region. This finding thus provides further support for the mechanism of site rearrangement and primary stress relaxation suggested in the defect model for the MMIE.

In this way, this paper can be regarded as an important step toward a complete understanding the physics underlying the MMIE.

## REFERENCES

1. V. Belostotsky, J. Non-Cryst. Solids 353 (2007) 1078.

2. N.H. Ray, J. Non-Cryst. Solids 15 (1974) 423.

3. A. Varshneya, Fundamentals of Inorganic Glasses, Academic Press, 1994, p. 99.

4. C.A. Angell, Chem. Rev. 90 (1990) 523.

5. M.J. Goldstain, J. Chem.Phys. 51 (1969) 3728.

6. M. Cohen and D. Turnbull, J. Chem. Phys. 31 (1959) 1164.

7. D. Turnbull and M. Cohen, J. Chem. Phys. 34 (1961) 120.

8. D. Turnbull and M. Cohen, J. Chem. Phys. 52 (1970) 3038.

9. P. Manns and R. Brückner, Glastechn. Ber. 56 (1983) 155.

10. R. Brückner and Y. Yue, J. Non-Cryst. Solids 175 (1994) 118.